\documentstyle[preprint,aps,epsfig,amsbsy,floats]{revtex}
 \tightenlines

\begin{document}
\newcommand{\beq}{\begin{equation}}
\newcommand{\eeq}{\end{equation}}
\newcommand{\beqa}{\begin{eqnarray}}
\newcommand{\eeqa}{\end{eqnarray}}
\newcommand{\sr}{\sqrt}
\newcommand{\fr}{\frac}
\newcommand{\mn}{\mu \nu}
\newcommand{\G}{\Gamma}

\draft \preprint{hep-th/0308191,~INJE-TP-03-08}
\title{ Logarithmic corrections to the Bekenstein-Hawking entropy for five-dimensional
black holes and de Sitter spaces}
\author{  Y.S. Myung\footnote{E-mail address:
ysmyung@physics.inje.ac.kr}}
\address{
Relativity Research Center and School of Computer Aided Science, Inje University,
Gimhae 621-749, Korea}
\maketitle

\begin{abstract}We calculate corrections to the Bekenstein-Hawking
entropy formula for the five-dimensional topological AdS
(TAdS)-black holes and topological de Sitter (TdS) spaces due to
thermal fluctuations. We can derive all thermal properties of  the
TdS spaces from those of the TAdS black holes by replacing $k$ by
$-k$. Also we obtain the same correction to  the Cardy-Verlinde
formula for TAdS and TdS cases including the cosmological horizon
of the Schwarzschild-de Sitter (SdS) black hole. Finally we
discuss the AdS/CFT and dS/CFT correspondences and their dynamic
correspondences.

\end{abstract}

\newpage
\section{Introduction}
Recently there are several works which show that for a large class
of black holes (AdS-Schwarzschild one), the Bekenstein-Hawking
entropy receives logarithmic corrections due to thermodynamic
fluctuations~\cite{KM1,KM2,CAL,GKS,BS}. The corrected formula
takes the form \beq \label{CEN} S=S_0-\fr{1}{2} \ln C_v + \cdots,
\eeq where $C_v$ is the specific heat of the given system at
constant volume and $S_0$ denotes the uncorrected
Bekenstein-Hawking entropy. Here an important point is that for
Eq.(\ref{CEN}) to make sense, $C_v$ should be positive. However,
the d-dimensional Schwarzschild black hole which is asymptotically
flat has a negative specific heat of
$C_v^{Sch}=-(d-2)S_0$~\cite{DMB}. This means that the
Schwarzschild black hole is never in thermal equilibrium and it
evaporates according to the Hawking radiation. But the
Schwarzschild black hole could be thermal equilibrium with a
radiation in a bounded box.  This is because the black hole has a
negative specific heat while the radiation has a positive one. The
two will be in thermal equilibrium if the box is bounded. The
AdS-Schwarzschild black hole  belongs to this category. On the
contrary, if the box is unbounded as the Schwarzschild black hole,
the black hole evaporates completely. Furthermore we note that a
cosmological horizon in five-dimensional de Sitter space has a
positive specific heat of $C_v^{dS}=3S_0$. Hence we can use the
Eq.(\ref{CEN}) to calculate the correction to the
Bekenstein-Hawking entropy of the cosmological horizon.

In this work, we find new  five-dimensional AdS-black holes and de
Sitter spaces which  give us positive specific heats and thus
logarithmic corrections to the entropy are achieved. These are the
topological AdS-black holes and topological de Sitter spaces. For
completeness, we study  thermal properties of the Schwarzschild-de
Sitter black hole.  Further we make corrections to the
Cardy-Verlinde formula which is a higher-dimensional version of
the two-dimensional Cardy formula. This formula realizes  the
holography principle through the A(dS)/CFT correspondences.

\section{Topological AdS black holes}
It is believed that black holes in asymptotically flat spacetime
should have spherical horizon. When introducing  a negative
cosmological constant, a black hole can have non-spherical
horizon. We call this the topological black hole\cite{BIR}. The
topological AdS black holes in five-dimensional spacetimes are
given by
 \beq ds^{2}_{TAdS}=
 -h(r)dt^2 +\fr{1}{h(r)}dr^2 +r^2 \left[d\chi^2
+f_{k}(\chi)^2(d\theta^2+ \sin^2 \theta d\phi^2) \right],
\label{BMT} \eeq where $k$ describes the horizon geometry with a
constant curvature. $h(r)$ and $f_k(\chi)$ are given by \beq
h(r)=k-\fr{m}{r^2}+ \fr{ r^2}{\ell^2},~~~ f_{0}(\chi) =\chi,
~f_{1}(\chi) =\sin \chi, ~f_{-1}(\chi) =\sinh \chi. \eeq

Here we define $k$=1,0, and $-1$ cases as the Schwarzschild-AdS
(SAdS) black hole~\cite{MP}, flat-AdS (FAdS) black hole, and
hyperbolic-AdS (HAdS) black hole~\cite{CAI1}, respectively. In the
case of $k=1,m=0$, we have an exact AdS$_5$-space with its
curvature radius $\ell$. However, $m \not=0$ generates the
topological AdS black holes. The only event horizon is given by
 \beq \label{EH} r_{EH}^2=
\fr{\ell^2}{2}\Big(-k+\sqrt{k^2 +4 m/\ell^2}\Big). \eeq For $k=1$,
we have both a small black hole ($r_{EH}^2<<\ell^2,4m<<\ell^2$)
 with  the horizon at $r=r_{EH}$, where $r_{EH}^2
 \simeq m$ and a large black hole  ($r_{EH}^2>>\ell^2,4m>>\ell^2 )$  with  the
horizon at $r=r_{EH}$ given by $r_{EH}^2 \simeq \sqrt{m}\ell$. For
$k=0$ case, one has the event horizon  at $r=r_{EH}$, where
$r_{EH}^2 = \sqrt{m}\ell$. In the case of $k=-1$, for $4m<<\ell^2$
one has the event horizon at $r=r_{EH}$, where $r_{EH}^2 \simeq
\ell^2+ m$ and for $4m>>\ell^2$ one has the  event horizon at
$r=r_{EH}$ given by $r_{EH}^2 \simeq \sqrt{m}\ell$. That is, one
always finds $r_{EH}^2>\ell^2$ for $k=-1$. This analysis is useful
to justify whether the corresponding specific heat is or not
positive.

The relevant thermodynamic quantities: reduced mass ($m$), free
energy ($F$), Bekenstein-Hawking entropy ($S_0$), Hawking
temperature ($T_H$), and  energy (ADM mass :$E=M$)  are given
by~\cite{LNOO} \beqa \label{TQ} &&m=r_{EH}^2
\Big(\fr{r_{EH}^2}{\ell^2}+ k \Big),~~F=-\fr{V_3 r_{EH}^2}{16 \pi
G_5}\Big(\fr{r_{EH}^2}{\ell^2}- k \Big),~~
S_0=\fr{V_3r_{EH}^3}{4G_5},~~\\ \nonumber &&T_H=\fr{k}{2 \pi
r_{EH}} +\fr{r_{EH}}{\pi \ell^2}~~E=F+T_H S_0 =\fr{3V_3m}{16 \pi
G_5}=M, \eeqa where $V_3$ is the volume of unit three-dimensional
hypersurface and $G_5$ is the five-dimensional Newton constant.
Using the above together with $C_v=(dE/dT)_V$, one finds \beq
\label{Heat1}C_v= 3 \fr{2r_{EH}^2+k\ell^2}{2r_{EH}^2-k\ell^2}S_0.
\eeq Here we obtain two positive specific heats for HAdS and FAdS
black holes \beq \label{SOH} C_v^{HAdS}>0,~~C_v^{FAdS}=3S_0>0,~~
{\rm for~ any~} r_{EH}. \eeq On the other hand one finds a
condition for positive specific heat for SAdS black
hole~\cite{MP}\beq \label{Heat2} C_v^{SAdS}>0,~~ {\rm for}~
r^2_{EH}>\ell^2/2. \eeq In the limit of $\ell\to \infty$, we
recover the negative specific heat ($C_v^{Sch}=-3S_s$) of the
Schwarzschild black hole. On the other hand, in the limit of
$\ell\to 0$  one finds a positive value of $C_v^{\ell \to 0}=3S_0$
for the large SAdS-black hole.

\section{Schwarzschild de Sitter black hole}
In order to find thermal property of a black hole in de Sitter
space,  we consider Schwarzschild de Sitter (SdS) black hole in
five-dimensional spacetimes~\cite{CAI2}  \beq ds^{2}_{SdS}=
-h(r)dt^2 +\fr{1}{h(r)}dr^2 +r^2 \left[d\chi^2
+\sin^2\chi(d\theta^2+ \sin^2 \theta d\phi^2) \right] \label{SDS}
\eeq where  $h(r)$ is given by \beq h(r)=1-\fr{m}{r^2}- \fr{
r^2}{\ell^2}.\eeq
 In
the case of $m=0$, we have an exact de Sitter space with its
curvature radius $\ell$. However, $m \not=0$ generates the SdS
black hole.  Here we have  two horizons.   The cosmological and
event horizons are given by \beq \label{3EH} r_{CH/EH}^2=
\fr{\ell^2}{2}\Big(1\pm \sqrt{1 -4 m/\ell^2}\Big). \eeq We
classify three cases : 1) $4m=\ell^2$, 2) $4m>\ell^2$, 3)
$4m<\ell^2$. The case of $4m=\ell^2$ corresponds to the maximum
black hole and the minimum cosmological horizon in asymptotically
de Sitter space (that is, Nariai black hole). In this case we have
$r_{EH}^2=r_{CH}^2=\ell^2/2=2m$. The case of $4m>\ell^2$ is not
allowed for the black hole in de Sitter space. The case of
$4m<\ell^2$ corresponds to a small black hole within the
cosmological horizon. In this case we have the cosmological
horizon at $r=r_{CH}$, where $r_{CH}^2 \simeq \ell^2-m$ and the
event horizon at $r=r_{EH}$ given by $r_{EH}^2 \simeq m$. Hence we
have two relations for the SdS solution: \beq \label{2Ineq} m \le
r^2_{EH} \le \ell^2/2,~~ \ell^2/2 \le r^2_{CH} \le \ell^2-m \eeq
which means that as $m$ increases from a small value to the
maximum  of $m=\ell^2/4$, a small black hole increases up to the
Nariai black hole. On the other hand the cosmological horizon
decreases from the maximum of $(\ell^2-m)$ to the minimum of
$\ell^2/2$.

 The
relevant thermodynamic quantities for two horizons   are given
by~\cite{CM,NOO} \beqa \label{2TQ} &&m=r_{EH/CH}^2
\Big(-\fr{r_{EH/CH}^2}{\ell^2}+ 1 \Big),~~F_{EH/CH}=\pm\fr{V_3
r_{EH/CH}^2}{16 \pi G_5}\Big(\fr{r_{EH}^2}{\ell^2}+1 \Big),~~
S_0=\fr{V_3r_{EH/CH}^3}{4G_5},~~\\ \nonumber
&&T_H^{EH/CH}=\pm\fr{1}{2 \pi r_{EH/CH}} \mp \fr{r_{EH/CH}}{\pi
\ell^2},~~E=F_{EH/CH}+T_H^{EH/CH} S_0 =\pm \fr{3V_3m}{16 \pi G_5},
\eeqa where $V_3$ is the volume of unit three-dimensional sphere.
Using the above relations, one finds \beq
\label{2Heat1}C_v^{EH/CH}= 3
\fr{2r_{EH/CH}^2-\ell^2}{2r_{EH/CH}^2+\ell^2}S_0. \eeq  Making use
of Eq.(\ref{2Ineq}), one finds negative specific heat for the
event horizon of the SdS black hole (ESdS) and positive specific
heat for the cosmological horizon (CSdS) \beq \label{2SOH}
C_v^{ESdS} \le 0,~~C_v^{CSdS} \ge 0, {\rm for~ any~} r_{EH}. \eeq
This means that the cosmological horizon is thermodynamically
stable while the event horizon is unstable. The equality sign
(that is, zero specific heat) holds for the Nariai black hole. In
the limit of $\ell\to \infty$, we recover the negative specific
heat ($C_v^{Sch}=-3S_s$) of the Schwarzschild black hole. On the
other hand, in the limit of $\ell\to 0$  one finds a positive
value of $C_v^{dS}=3S_0$ for the exact de Sitter space.

\section{Topological de Sitter space}
The topological de Sitter (TdS) solution was originally introduced
to check the mass bound conjecture in de Sitter space: any
asymptotically de Sitter space with the mass greater than exact de
Sitter space has a cosmological singularity~\cite{TDS}.  For our
purpose, we consider the topological de Sitter solution in
five-dimensional spacetimes \beq ds^{2}_{TdS}= -h(r)dt^2
+\fr{1}{h(r)}dr^2 +r^2 \left[d\chi^2 +f_{k}(\chi)^2(d\theta^2+
\sin^2 \theta d\phi^2) \right], \label{Tds} \eeq where $k=0,\pm1$.
$h(r)$ and $f_k(\chi)$ are given by \beq h(r)=k+\fr{m}{r^2}- \fr{
r^2}{\ell^2},~~~ f_{0}(\chi) =\chi, ~f_{1}(\chi) =\sin \chi,
~f_{-1}(\chi) =\sinh \chi. \eeq Requiring $m>0$, the black hole
disappears and instead a naked singularity occurs at $r=0$. Here
we define $k=1,0,-1$ cases as the Schwarzschild-topological de
Sitter (STdS) space, flat-topological de Sitter (FTdS) space, and
hyperbolic--topological de Sitter (HTdS) space, respectively. In
the case of $k=1,m=0$, we have an exact de Sitter space with its
curvature radius $\ell$. However, $m>0$ generates the topological
de Sitter spaces. The only cosmological horizon exists as \beq
\label{CH} r_{CH}^2= \fr{\ell^2}{2}\Big(k+\sqrt{k^2 +4 m/\ell^2}
\Big). \eeq

 For $k=-1$ case we have both a small cosmological horizon ($r_{CH}^2<<\ell^2,4m<<\ell^2$)
 with  the horizon at $r=r_{CH}$, where $r_{CH}^2
 \simeq m$ and a large cosmological horizon ($r_{CH}^2>>\ell^2,4m>>\ell^2 )$  with  the
horizon at $r=r_{CH}$ given by $r_{CH}^2 \simeq \sqrt{m}\ell$. For
$k=0$ case, one has the cosmological horizon  at $r=r_{CH}$, where
$r_{CH}^2 = \sqrt{m}\ell$. In the case of $k=1$, for $4m<<\ell^2$
one has the  cosmological horizon at $r=r_{CH}$, where $r_{CH}^2
\simeq \ell^2+ m$ and for $4m>>\ell^2$, one has the cosmological
horizon at $r=r_{CH}$, where $r_{CH}^2 \simeq \sqrt{m}\ell$. Here
we have $r_{CH}^2>\ell^2$ for $k=1$ case. This analysis is useful
to justify whether the  specific heat of the cosmological horizon
is or not positive.

The relevant thermodynamic quantities for the cosmological horizon
are calculated as~\cite{CAI2}\beqa \label{3TQ} &&m=r_{CH}^2
\Big(\fr{r_{CH}^2}{\ell^2}- k \Big),~~F=-\fr{V_3 r_{CH}^2}{16 \pi
G_5}\Big(\fr{r_{CH}^2}{\ell^2}+ k \Big),~~
S_0=\fr{V_3r_{CH}^3}{4G_5},~~\\ \nonumber &&T_H=-\fr{k}{2 \pi
r_{CH}} +\fr{r_{CH}}{\pi \ell^2},~~E=F+T_HS =\fr{3V_3m}{16 \pi
G_5}=M, \eeqa where $V_3$ is the volume of unit three-dimensional
hypersurface. Using the above relations, one finds \beq
\label{3Heat1}C_v= 3 \fr{2r_{CH}^2-k\ell^2}{2r_{CH}^2+k\ell^2}S_0.
\eeq Here we have positive specific heats for STdS and FTdS spaces
\beq \label{3SOH} C_v^{STdS}>0,~~C_v^{FTdS}=3S_0>0, ~{\rm for~
any~} r_{CH}. \eeq On the other hand  one finds the positive
specific heat for HTdS space, \beq \label{3Heat2} C_v^{HTdS}>0
~~{\rm when ~} r^2_{CH}>\ell^2/2. \eeq We note that all results of
the TdS solution can be recovered from the TAdS solution by
replacing $k$ by $-k$. This relation will play an important role
in understanding de Sitter space  in terms of AdS solution.

\section{correction to entropy and Cardy-Verlinde formula}
In this section we make corrections to the Bekenstein-Hawking
entropy according to the formula of Eq.(\ref{CEN}). For the FAdS
black hole and FTdS solution one finds $C_v=3S_0$ without any
approximation. However, other cases (HAdS and SAdS black holes,
CSdS, STdS and HTdS spaces) lead to $C_v \simeq 3S_0$ when
choosing large black holes ($r_{EH}^2>>\ell^2$) and large
cosmological horizons ($r_{CH}^2>>\ell^2$). As far as $C_v \simeq
3S_0$ is guaranteed, the logarithmic correction to the
Bekenstein-Hawking entropy is given by \beq \label{CENT}
S^{TAdS,CSdS,TdS}=S_0-\fr{1}{2} \ln S_0 + \cdots. \eeq Note that
there is no correction to the SdS black hole horizon (ESdS):
$S_{EH}^{ESdS}=S_{0}^{ESdS}$. Thus we do not consider this case
for correction.

The holographic principle means that the number of degrees of
freedom associated with the bulk gravitational dynamics is
determined by its boundary spacetime. The AdS/CFT correspondence
represents a realization of this principle~\cite{HOL}. Further,
for a strongly coupled CFT with its AdS dual, one obtains the
Cardy-Verlinde formula~\cite{VER}. Indeed this formula holds for
various kinds of asymptotically AdS spacetimes including the TAdS
black holes~\cite{CAI1}. Also this formula holds for a few of
asymptotically de Sitter spacetimes including the SdS black hole
and TdS spacetimes~\cite{CAI2}. Hence it needs to correct the
Cardy-Verlinde formula if possible. For this purpose, we have to
define thermodynamic quantities described by the boundary CFT
through the A(dS)/CFT correspondences~\cite{witten}.  The relation
between the five-dimensional bulk and four-dimensional boundary
quantities is given by $E_4=(\ell/R)E,~ T=(\ell/R)T_H$ where $R$
satisfies $T>1/R$ but one has the same entropy : $S_4=S_0$. We
note that the boundary physics is described by the CFT-radiation
matter with the equation of state: $p=E_4/3V_3$. Then a
logarithmic correction is being determined by the Casimir energy
defined by $E_c=3(E_4+pV_3-TS_0)$. We obtain \beqa &&E_c^{TAdS}=k
\fr{3\ell r_{EH}^2 V_3}{8
\pi G_5 R}+ \fr{3}{2} T \ln S_0,\\
 &&E_{c}^{CSdS}=-
\fr{3\ell
r_{CH}^2 V_3}{8 \pi G_5 R}+\fr{3}{2} T \ln S_0,\\
 &&E_c^{TdS}=-k \fr{3\ell r_{CH}^2 V_3}{8 \pi G_5 R}+ \fr{3}{2} T
\ln S_0. \eeqa

 Substituting this expression into the
Cardy-Verlinde formula, one finds  \beqa && TAdS~ :~~\fr{ 2 \pi
R}{3 \sqrt{|k|}} \sqrt{|E_c(2E_4-E_c)|} \simeq S_0 + \fr{\pi R
\ell T}{2kr_{EH}^3}
\Big( \fr{r_{EH}^4}{\ell^2}-kr_{EH}^2 \Big) \ln S_0, \\
&& CSdS~:~~\fr{ 2 \pi R}{3} \sqrt{|E_c(2E_4-E_c)|} \simeq S_0 -
\fr{\pi R \ell T}{2r_{CH}^3}
\Big( \fr{r_{CH}^4}{\ell^2}+r_{CH}^2 \Big) \ln S_0, \\
&& TdS~:~~\frac{ 2 \pi R}{3 \sqrt{|k|}} \sqrt{|E_c(2E_4-E_c)|}
\simeq S_0 - \fr{\pi R \ell T}{2kr_{CH}^3} \Big(
\fr{r_{CH}^4}{\ell^2}+kr_{CH}^2 \Big) \ln S_0.  \eeqa All
coefficients in front of  $\ln S_0$ in the above equations are
transformed as~\cite{LNOO,NOO} \beqa && \fr{\pi R\ell T}{2 k
r_{EH}^3} \Big(
\fr{r^4_{EH}}{\ell^2} -kr_{EH}^2\Big)=\fr{(4E_4-E_c)(E_4-E_c)}{2(2E_4-E_c)E_c}, \\
&& -\fr{\pi R\ell T}{2  r_{CH}^3 }\Big(
\fr{r^4_{CH}}{\ell^2} +r_{CH}^2\Big)=\fr{(4E_4-E_c)(E_4-E_c)}{2(2E_4-E_c)E_c}, \\
&& -\fr{\pi R\ell T}{2 k r_{CH}^3 }\Big( \fr{r^4_{CH}}{\ell^2}
+kr_{CH}^2\Big)=\fr{(4E_4-E_c)(E_4-E_c)}{2(2E_4-E_c)E_c}. \eeqa

Finally we obtain the same corrected formula for the
Cardy-Verlinde formula as \beq \label{CCV}S^{TAdS,CSdS,TdS}_{CV}
\simeq \fr{ 2 \pi R}{3 \sqrt{|k|}} \sqrt{|E_c(2E_4-E_c)|}-
\fr{(4E_4-3E_c)E_4}{2(2E_4-E_c)E_c} \ln \Big(\fr{ 2 \pi R}{3
\sqrt{|k|}} \sqrt{|E_c(2E_4-E_c)|}\Big).\eeq

\section{Discussion}
\begin{table}
 \caption{Summary of specific heats, boundary CFT energy and uncorrected Casimir energy
 for 5D TAdS black holes, TdS spaces and SdS black hole.}
 \begin{tabular}{lp{4.5cm}p{3cm}}
 thermodynamical system   & $C_v$ & $E_4(E_c)$ \\ \hline
  HAdS & + & +($-$) \\
FAdS & +($3S_0$) & +($0$) \\
SAdS & + if $r^2_{EH}>
\ell^2/2$ & +($+$) \\
STdS & + & +($-$) \\
FTdS & + ($3S_0$) & +($0$) \\
HTdS & + if $ r^2_{CH}>
\ell^2/2$& +($+$) \\
ESdS & $-$ & +($+$) \\
CSdS & + & $-(-$)
 \end{tabular}
 \end{table}

First of all we summarize our result. As is shown in TABLE I,
$C_v=3S_0$ for FAdS and FTdS cases without any approximation. Also
we have $C_v>0$ for HAdS, STdS and CSdS cases whereas $C_v>0$ if
$r_{EH/CH}^2>\ell^2/2$ for  SAdS black holes and HTdS space. Note
that $C_v^{ESdS}<0$ for the black hole in de Sitter space.
However, choosing large black holes  and large de Sitter spaces
($C_v \simeq 3S_0$) except ESdS case leads to the same corrected
formulae for the Bekenstein-Hawking entropy Eq.(\ref{CENT}) and
the Cardy-Verlinde formula Eq.(\ref{CCV})\footnote{Also  a
similarly corrected Cardy-Verlinde formula for the TNRdS space
appeared in\cite{SET}.}. Concerning the A(dS)/CFT correspondences,
we remind that the boundary CFT energy ($E_4$) should be positive
in order for it to make sense. However, one
 finds from TABLE that $E_4^{CSdS}<0$ for the cosmological horizon of the SdS black
 hole. It suggests that the dS/CFT correspondence is not valid for this case.
 Also the Casimir energy ($E_c$) is related to the central charge
 of the corresponding CFT. Hence if it is negative, one  may obtain the
 non-unitary CFT. In this sense, HAdS, STdS, and CSdS cases are
 problematic. Further we comment on the extension of this approach
 to the dynamic A(dS)/CFT correspondence by introducing a moving
 domain wall in the bulk background (brane world cosmology). Although there is no problem in
   the AdS-back hole background~\cite{SV,MYU1}, there remains  problem in
   interpreting the cosmic energy density
   in compared with the static energy like $E_4$ in the de Sitter background~\cite{MYU2}.

Finally we wish to mention that through this work, we can derive
all thermal properties of  the topological de Sitter (TdS) spaces
from the topological anti-de Sitter (TAdS) black holes by
replacing $k$ by $-k$.

\section*{Acknowledgments}

This work was supported in part by KOSEF, Project No.
R02-2002-000-00028-0.

\end{document}